\documentclass{elsarticle}
\usepackage{amsmath,amssymb}
\usepackage{lscape}

\begin{document}
\title{Results of a Search for Paraphotons with Intense X--ray Beams at SPring--8}
\author[phys]{T.~Inada}
\ead{tinada@icepp.s.u-tokyo.ac.jp}
\author[icepp]{T.~Namba}
\author[phys]{S.~Asai}
\author[icepp]{T.~Kobayashi}
\author[riken]{\\Y.~Tanaka}
\author[riken]{K.~Tamasaku}
\author[riken]{K.~Sawada}
\author[riken]{T.~Ishikawa}
\address[phys]{Department of Physics, Graduate School of Science, the University of Tokyo, 7--3--1 Hongo, Bunkyo, Tokyo 113--0033, Japan}
\address[icepp]{International Center for Elementary Particle Physics, the University of Tokyo, 7--3--1 Hongo, Bunkyo, Tokyo 113--0033, Japan}
\address[riken]{Spring-8/RIKEN Harima Institute, 1--1--1 Kouto, Sayo-cho, Sayo-gun, Hyogo 679--5148, Japan}

\begin{abstract}
A search for paraphotons, or hidden $U(1)$ gauge bosons, is performed using
an intense X--ray beamline at SPring--8.
``Light Shining through a Wall'' technique is used in this
search.
No excess of events above background is observed.
A stringent constraint is obtained on the photon--paraphoton mixing angle,
$\chi < 8.06\times 10^{-5}\ (95\%\ {\rm C.L.})$ for $0.04\ {\rm eV}<m_{\gamma^{\prime}} < 26\ {\rm keV}$.
\end{abstract}

\maketitle

\section{Introduction}
Paraphotons, or hidden sector photons are gauge bosons of
hypothetical $U(1)$ symmetry.
Many extensions of the Standard Model
predict such a symmetry~\cite{cite_theory}.
Some of them also predict tiny mixing of paraphotons with ordinary
photons through very massive particles
which have both electric and hidden charge~\cite{cite_couple}.
This effective mixing term induces flavor oscillations between paraphotons and
ordinary photons~\cite{cite_osc}.
With this oscillation mechanism, a high sensitive search can be
done with a method called ``Light Shining through a
Wall (LSW)'' technique\cite{cite_prop}, in which incident photons oscillate into paraphotons that
are able to pass through a wall and oscillate back into photons.

Recently, a detailed theoretical calculation has been performed for axion LSW experiment\cite{cite_highmass}.
Since both axion-- and paraphoton--conversion are described as the same quantum oscillations,
the conversion probability for axions can be interpreted as that of paraphotons by replacing parameters from $\frac{\beta \omega}{m^{2}}$ to $\chi$ in Eq.~(29) in \cite{cite_highmass}.
After propagation in vacuum for length $L$, the probability of
converting a paraphoton into a photon (or vice versa) is given by,
\begin{equation}
p_{\gamma \leftrightarrow \gamma'}\left(L\right) =
\left(\frac{\omega + \sqrt{\omega^{2}-m_{\gamma^{\prime}}^{2}}}{\sqrt{\omega^{2}-m_{\gamma^{\prime}}^{2}}}\chi\right)^{2}
\sin^{2}\left( \frac{L}{2} \left(\omega-\sqrt{\omega^{2}-m_{\gamma^{\prime}}^{2}}\right) \right),
\label{eq:prob}
\end{equation}
where $\chi$ is the mixing angle, $m_{\gamma '}$ is the mass of the
paraphoton, and $\omega$ is the energy of photon.
For low mass region ($m_{\gamma^{\prime}}\ll\omega$),
it becomes a well-known expression of a neutrino-like oscillation;
$p_{\gamma \leftrightarrow \gamma '}(L)=4\chi^{2}\sin^{2}(m_{\gamma '}^{2} L/4\omega)$.

Searches have been performed with this LSW technique
by using optical photons~\cite{cite_optical} or
microwave photons~\cite{cite_rf}, without any evidence.
Useful summary papers are available (see {\it e.g.} \cite{cite_ringwald}).
For an axion-LSW search, an experiment using X--rays has been performed at ESRF\cite{cite_esrf}.

In this letter, we report a new search for paraphotons with the LSW method.
We use an intense X--ray beam created by a long undulator at SPring--8
synchrotron radiation facility to search paraphotons
whose mass is in the ($10^{-1}$--$10^{4}$) eV region.

\section{Experimental Setup}

\begin{figure}
\begin{center}
\includegraphics[angle=-90, width=120mm]{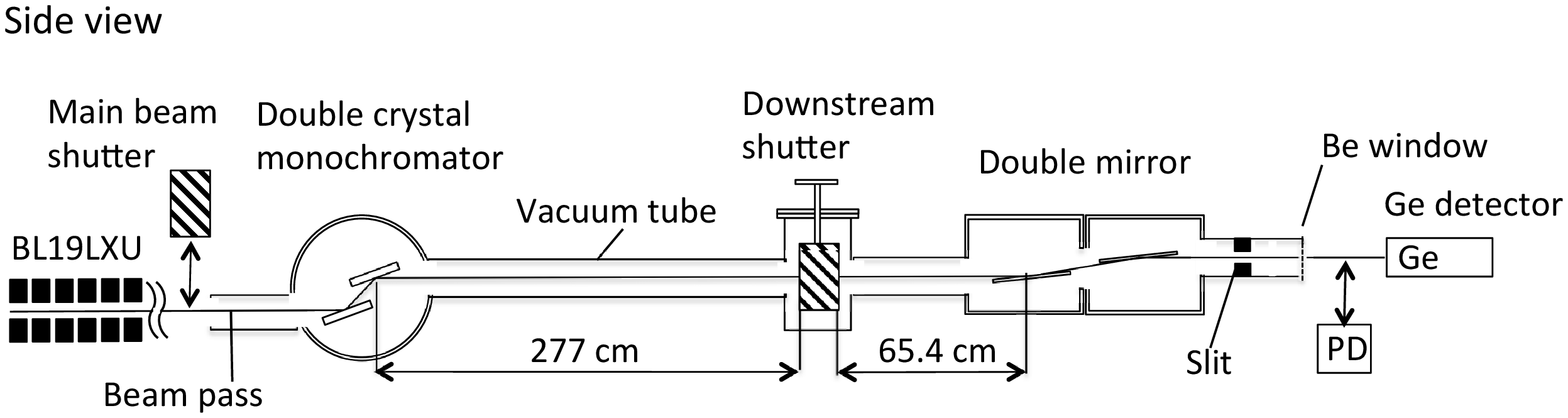}
\end{center}
\caption{Schematic view of our experimental setup.}
\label{fig:beamline}
\end{figure}

BL19LXU~\cite{cite_bl} beamline
at SPring--8 (Fig.~\ref{fig:beamline}) is used for X--ray source.
A 30-m long undulator is placed
on the electron storage ring as shown in Fig.~\ref{fig:beamline}.
A bunch length of electrons in the storage ring is $40$ ps,
and a bunch interval is $23.6$ ns.
Structure of a X--ray beam represents the bunch structure of electrons,
but we regard it as a
continuous beam because time resolution of X--ray detector is larger than
this structure.
An energy of the X--ray beam is tunable between $7.2$ and $18$ keV
by changing a gap width of the undulator.
Higher energy of its 3rd harmonics ($21.6\sim 51$ keV) is also available.
X--ray beam is monochromated with a ${\rm Si}(111)$
double crystal monochromator to the level
of $\Delta \omega/\omega \sim 10^{-4}$.
A reflection angle is determined from Bragg condition, and is typically $\sim$ 100 mrad for energies we use.
A beam size is about $1 \ {\rm mm}$, and a vertical profile ($\rho (x)$)
is measured with a slit with $10 \ {\rm \mu} {\rm m}$ pitch.
Shape of $\rho (x)$ is similar to Gaussian whose FWHM
is $383 \ {\rm \mu} {\rm m}$.

From the monochromator, the X--ray beam is guided through vacuum tubes,
whose length is about $3.5$ m.
Tubes are evacuated better than $4\times 10^{-5}\ {\rm Pa}$,
and a double mirror is placed at the downstream edge of the tube.
These mirrors are adjusted for the total reflection,
and their reflection angle is tuned at 3.0 mrad (or 2.0 mrad) during
our search (only at 26 keV search).
They serve as a beam-pass filter, since only X--ray beams satisfying a severe condition of total reflection are bounced up and the other off-axis background photons are blocked.
The X--ray beam changes its path with these mirrors and only the
reflected beam is selected with a slit, and guided to the X--ray
detector.

Two beam shutters are placed in the beamline.
Main Beam Shutter (MBS) is placed just before the monochromator,
and DownStream Shutter (DSS) is placed between
the monochromator and the mirrors.
Photon changes into paraphoton in a vacuum tube between the monochromator and DSS,
and then changes back inversely in the region between DSS and the mirrors.
Each length at the beam center is $(277\pm 2)$ cm and $(65.4\pm 0.5)$ cm,
respectively\let\thefootnote\relax\footnotetext{In this paper, all error values represent 1 sigma.}.

A germanium detector (Canberra BE2825) is used to detect X--ray signal.
A diameter and thickness of its crystal is 60 mm and 25 mm, respectively.
Signal of Ge detector is shaped with an amplifier (ORTEC 572) and recorded by a peak hold ADC (HOSHIN C-011).
Energy resolution of the detector is measured with
${}^{55}{\rm Fe}$, ${}^{68}{\rm Ge}$, ${}^{57}{\rm Co}$,
and ${}^{241}{\rm Am}$ sources,
and typical energy resolution at 10 keV is 0.17 keV ($\sigma$: standard deviation).
Absolute efficiencies of the X--ray detector ($\epsilon$) are
also measured by the same sources.
Measured efficiencies are consistent with GEANT4 Monte Carlo results,
which includes all attenuations in the air, carbon composite window (thickness$=600$ $\mu$m) of the detector, and surface dead layer (thickness $=\left( 7.7\pm 0.9 \right) $ $\mu$m) of the germanium crystal.

The detector is shielded by lead blocks whose thickness is about 50 mm
except for a collimator on the beam axis whose hole diameter is $30$ mm,
much larger than the X--ray beam size.
The position of the collimator and the germanium crystal against the beam
is adjusted by using a photosensitive paper which is sensitive to the X--ray.

After the monochromator reaches thermal equilibrium,
beam flux becomes stable.
Absolute flux of the X--ray beam and its stability are monitored by a silicon PIN
photodiode (Hamamatsu S3590-09, $\textrm{thickness} = 300$ $\mu$m).
This photodiode is inserted
in front of the collimator of the lead shield, and DSS is opened
for the flux measurement.
During this measurement, the collimator hole is closed to
avoid the radiation damage to the germanium detector.
The energy deposited on the PIN photodiode is calculated
using its output current and the W-value of silicon (${\rm W}=3.66\ \rm{eV}$).
Fraction of the X--ray energy deposition in the PIN photodiode is computed
with GEANT4 simulation for each energy.
To correct the saturation effect of the PIN photodiode,
thin aluminum foils are inserted before the photodiode to attenuate X--ray flux.
Attenuation coefficient of aluminum is also checked by GEANT4 simulation.
The flux can be measured with an accuracy of less than 5\%.

\section{Measurement and Analysis}

A paraphoton search is performed from 14th to 20th June, 2012.
9 measurements are performed with different X--ray energies from $7.27$ keV to $26.00$ keV.
Results are summarized in Tab.~\ref{tab:result}.
Beam intensities ($I$) are monitored every 3--4 hours
by the PIN photodiode as described in the previous section.
Time drifts of the beam flux $(<10\%)$ are confirmed only at the beginning of the measurement since, due to heavy heat load, it takes about 30 minutes for experimental setup to become thermally stable.
Fluxes which get well-stabilized in the thermal equilibrium are listed and used for the analysis.
Energy calibration of the detector is also performed every 3--4 hours with a $^{57}$Co source.

\begin{figure}
\hspace{-25mm}
\begin{center}
\includegraphics[angle=-90, width=0.9\textwidth]{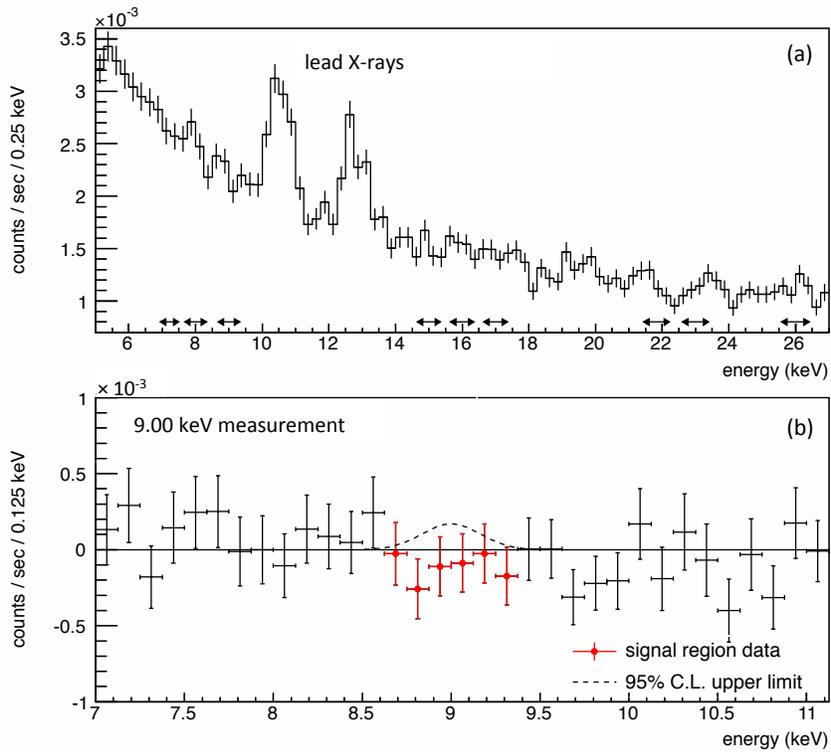}
\end{center}
\caption{(a) Energy spectrum obtained with MBS closed (BG spectrum).
Arrows show regions in which paraphoton searches are performed.
(b) Energy spectrum measured at $\omega =$ 9.00 keV.
Background contributions are subtracted using the spectrum (a).
Signals with statistical errors are shown in cross, and dotted line shows
obtained upper limit (95\% C.L.) of the signal.}
\label{fig:spectrum}
\end{figure}

BG spectrum (Fig.~\ref{fig:spectrum} (a)) is measured
from 16th to 17th June with MBS closed.
The other setup including the lead shields are completely the same as in the
paraphoton searches.
Total livetime of BG measurement is $1.6\times10^{5}$ s.
The BG rate at 7.00 keV is $(10.9\pm0.3)\times10^{-3}$ ${\rm s^{-1}\ keV^{-1}}$
and gradually decreases toward $(4.6\pm0.2)\times10^{-3}$ ${\rm s^{-1}\ keV^{-1}}$ at 26.00 keV.
No apparent structure is observed in the measured BG spectrum except for 10.6 keV and 12.6 keV, X--rays from the lead shields.

We define signal region as inside $\pm 2\sigma$
around the beam energy $\omega$.
Since signal regions are not overlapped among all measurements,
the BG spectrum is commonly used for all subtractions (Fig.~\ref{fig:spectrum} (b)).
The subtracted signal rates ($\Delta N$) are also shown
in Tab.~\ref{tab:result}, and no significant excess is observed for all 9 measurements.
Using these rates, we set upper limits on signal rates of measurements.
Gaussian distributions are assumed from center values and
the standard deviations of $\Delta N$, and $95\%$ C.L. positions in the
physical (i.e. positive) regions are set as a signal upper limit ($\Delta N_{95}$).
Finally, the upper limits on the LSW probability ($P_{95}$) are
obtained by $\Delta N_{{\rm 95}}/\epsilon I$.

\begin{figure}
\hspace{-40mm}
\begin{center}
\includegraphics[angle=0, width=1.\textwidth]{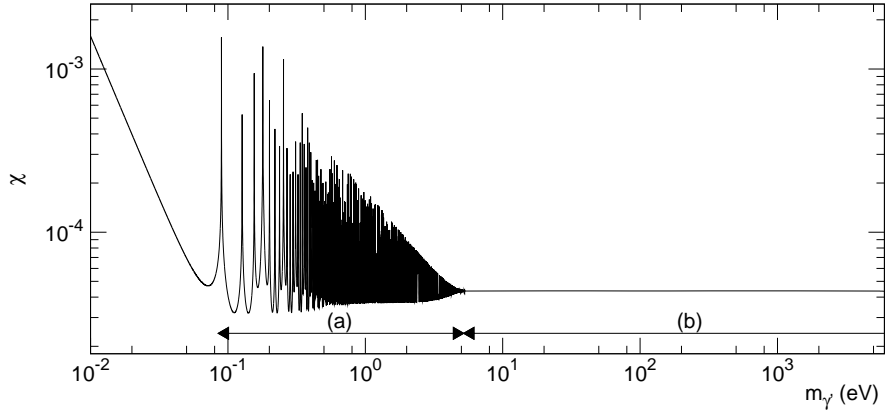}
\end{center}
\caption{Upper limit (95\% C.L.) on $\chi$ as a function of paraphoton mass ($m_{\gamma^{\prime}}$) obtained with only one search at $\omega =$ 9.00 keV.
Spiky structure is due to the photon--paraphoton oscillation as shown in Formula (\ref{eq:prob}).
Spikes are smeared for heavier mass region (labeled as ``(b)''), because of the smearing effect in Formula (\ref{eq:lsw}).}
\label{fig:limit_9kev}
\end{figure}

To translate $P_{95}$ to the limit on the mixing parameter $\chi$,
we need to consider $\rho (x)$ of the X--ray beam.
Since the incident angles of the beam into
the second crystal of the monochromator and the first mirror
are very shallow,
$\rho (x)$ affects the lengths of the oscillation regions.
As a result, these lengths are smeared by $\rho (x)$, and
the LSW probability is written as,
\begin{equation}
P=
\int _x \rho (x)\,
p_{\gamma \to \gamma '}\left(L_{1}\left(x\right)\right)\,
p_{\gamma ' \to \gamma}\left(L_{2}\left(x\right)\right)\,
{\rm d}x.
\label{eq:lsw}
\end{equation}
Here, $L_1(x)$ is the length of photon $\to$ paraphoton oscillation region
modified by the vertical position, and $L_2(x)$ is that of the re-oscillation
region.
The integration is numerically calculated for each $\omega$ as a function
of $m _{\gamma '}$, and $P_{95}$ is translated to the limit on $\chi$.
Figure \ref{fig:limit_9kev} shows 95\% C.L. limit obtained using a data set of 9.00 keV measurement, and upper side of the line is excluded.
The limit is smoothed by the smearing effect of $\rho(x)$ and becomes constant for masses
from 5 eV up to around 9 keV (labeled as ``(b)").

\begin{figure}
\hspace{-25mm}
\begin{center}
\includegraphics[width=0.9\textwidth]{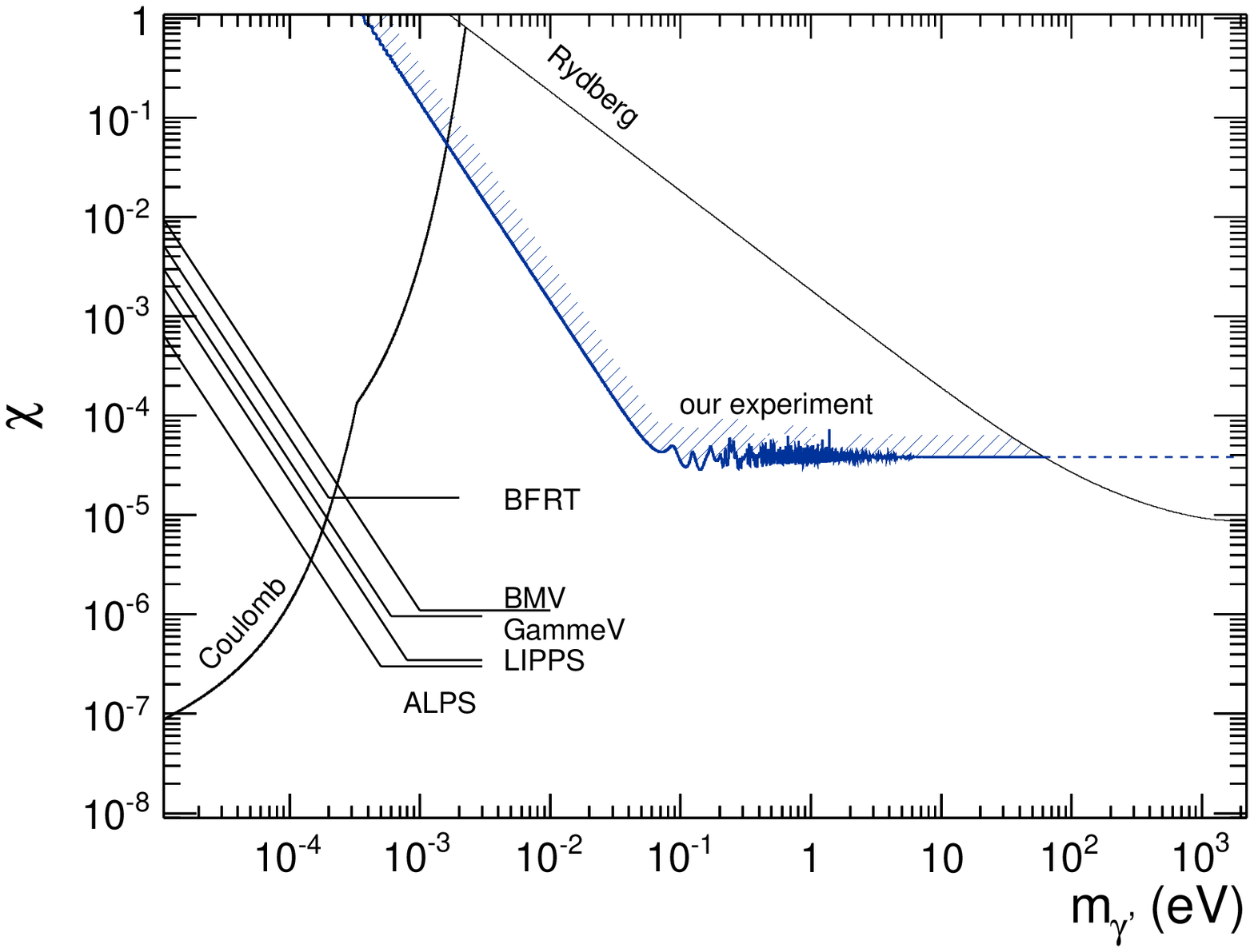}
\end{center}
\caption{Obtained 95\% C.L. limit on the paraphoton mixing angle compared with other laboratory experiments.
Rydberg is a limit from the measurement of Rydberg
atoms~\cite{cite_rydberg},
Coulomb is from the Coulomb low confirmation~\cite{cite_coulomb},
and BFRT, BMV, GammeV, LIPPS, and ALPS are from LSW experiments using optical laser~\cite{cite_optical}.}
\label{fig:limit_comb}
\end{figure}

Limit oscillations in the region (a) are ruled out by the combination of 9 measurements.
Combined results are obtained by the described procedure using $\chi^{4}$ distributions and multiplying each others.
95\% C.L. upper limit of the combined result is shown in Fig.~\ref{fig:limit_comb} with other results.
The worst value of $\chi _{{\rm worst}} = 8.01\times 10^{-5}$ appears at 1.39 eV.
Systematic errors on energy scale of the detector and oscillation region lengths,
including contribution from the uncertainty of $\rho(x)$ ($\Delta L<0.5$ mm),
are estimated by varying the parameters and obtained to be $\Delta \chi_{{\rm worst}} / \chi_{{\rm worst}} = ^{+0.52}_{-0.15}\%$.
Conservatively, $\chi _{{\rm worst}}+\Delta \chi _{{\rm worst}}$ represents our final result,
\begin{equation}
\chi < 8.06 \times 10^{-5}\ \ {\rm (95\%\ C.L.).}
\end{equation}
This result is valid for masses up to 26 keV, the maximum beam energy of our search.
Our result is the most stringent for masses around eV region as a terrestrial search.

\section{Conclusion}
A paraphoton search is performed at BL19LXU beamline in SPring--8
synchrotron radiation facility.
A double oscillation process, ``photons oscillating into paraphotons
and oscillating back into photons'', is assumed, and photons passing
through a wall are searched. No such photons are observed, and
a new limit on the photon--paraphoton mixing angle,
$\chi < 8.06\times 10^{-5}$ (95\% C.L.) is obtained for $0.04\ {\rm eV}<m_{\gamma^{\prime}} < 26\ {\rm keV}$.

\section*{Acknowledgements}
The synchrotron radiation experiment is performed at BL19LXU in SPring--8
with the approval of RIKEN (Proposal No.~20120088).
Sincere gratitude is also expressed to Dr. Suehara and Mr. Ishida for useful discussions.
Work of T. Inada is supported in part by Advanced Leading Graduate Course for Photon Science (ALPS) at U. Tokyo.

\clearpage

\clearpage

\begin{landscape}
\begin{table}[t]
\begin{center}
\begin{tabular}{ccccccccc}
\hline
\hline
beam &
livetime &
detector &
event rate &
BG subtracted rate &
signal &
beam &
detector &
LSW prob.
\\

energy &
&
resolution &
in ($\omega\pm 2\sigma$) &
in ($\omega\pm 2\sigma$) &
upper limit &
flux &
efficiency &
upper limit
\\

$\omega$ (keV)&
($10^{4}$ s)&
$\sigma$ (keV)&
$N$ ($10^{-3}\ {\rm s^{-1}}$)&
$\Delta N$ ($10^{-4}\ {\rm s^{-1}}$)&
$\Delta N_{{\rm 95}}$ ($10^{-4}\ {\rm s^{-1}}$)&
$I$ ($10^{13}$ ${\rm s^{-1}}$)&
$\epsilon$ (\%)&
$P_{{\rm 95}}$ $(10^{-16})$
\\
\hline

\ \ 7.27&
2.5&
0.16&
$7.0\pm0.5$&
$-0.9\pm5.7$&
11.0&
7.6&
23&
0.63
\\

\ \ 8.00&
2.0&
0.16&
$6.5\pm0.6$&
$-3.8\pm6.1$&
10.3&
8.9&
33&
0.35
\\

\ \ 9.00&
3.2&
0.17&
$5.3\pm0.4$&
$-7.6\pm4.5$&
5.5&
8.3&
46&
0.14
\\

15.00&
1.9&
0.18&
$4.2\pm0.5$&
$-3.4\pm5.0$&
8.2&
4.6&
51&
0.35
\\

16.00&
2.1&
0.18&
$4.2\pm0.4$&
$-3.1\pm4.8$&
7.9&
3.7&
56&
0.38
\\

17.00&
2.5&
0.18&
$4.2\pm0.4$&
$-2.1\pm4.5$&
7.8&
2.3&
61&
0.56
\\

21.83&
2.5&
0.19&
$4.2\pm0.4$&
$+4.2\pm4.3$&
12.2&
0.72&
76&
2.2
\\

23.00&
2.0&
0.20&
$3.9\pm0.4$&
$+1.2\pm4.7$&
10.5&
0.43&
78&
3.1
\\

26.00&
2.6&
0.21&
$4.8\pm0.4$&
$+7.6\pm4.6$&
15.6&
1.3&
83&
1.4
\\
\hline
\hline

\end{tabular}
\end{center}
\caption{Summary of 9 measurements of the paraphoton search.
Errors are one standard deviation statistical errors.}
\label{tab:result}
\end{table}
\end{landscape}

\clearpage

\clearpage

\clearpage

\begin{thebibliography}{00}
\bibitem{cite_theory}{J.~Jaeckel and A.~Ringwald,
	Ann. Rev. Nucl. Part. Sci. 60(2010)405.}
\bibitem{cite_couple}{B.~Holdom, Phys. Lett. B 166(1986)196.}
\bibitem{cite_osc}{L.~B.~Okun, JETP 56(1982)502.}
\bibitem{cite_prop}{K. Van Bibber {\it et al.},
	Phys. Rev. Lett. 59(1987)759.}
\bibitem{cite_highmass}{S. L. Adler {\it et al.},
	Ann. Phys. 323(2008)2851.}
\bibitem{cite_optical}{
BFRT Collaboration, R. Cameron {\it et al.},
	Phys. Rev. D 47(1993)3707;

BMV Collaboration, M. Fouche {\it et al.},
	Phys. Rev. D 78(2008)032013;

GammeV Collaboration, A. Chou {\it et al.},
	Phys. Rev. Lett. 100(2008)080402;

LIPPS Collaboration, A. Afanasev {\it et al.},
	Phys. Lett. B 679(2009)317;

ALPS Collaboration, K. Ehret {\it et al.},
	Phys. Lett. B 689(2010)149.}
\bibitem{cite_rf}{
ADMX Collaboration, A. Wagner {\it et al.},
	Phys. Rev. Lett. 105(2010)171801;

M. Betz and F. Caspers,
	Conf. Proc. C 1205201(2012)3320.}
\bibitem{cite_ringwald}{A. Ringwald,
	Dark Universe 1(2012)116.}
\bibitem{cite_esrf}{R. Battesti {\it et al.},
	Phys. Rev. Lett. 105(2012)250405.}
\bibitem{cite_bl}{M. Yabashi {\it et al.},
	Nucl. Instrum. Meth. A 467-468(2001)678.}
\bibitem{cite_rydberg}{
R. G. Beausoleil {\it et al.},
	Phys. Rev. A 35(1987)4878.}
\bibitem{cite_coulomb}{E. R. Williams, J. E. Faller and H. A. Hill.
	Phys. Rev. Lett. 26(1971)721.}
\end{thebibliography}
\end{document}